\documentclass[10pt]{article}
\usepackage[margin=1in]{geometry}   
\usepackage{authblk}  

\usepackage{bbm}
\usepackage{makeidx}         
\usepackage{graphicx}        

\usepackage{cite}
\usepackage{amssymb,amsthm,bm,mathrsfs,mathtools,amsmath,url,color,extarrows}
\usepackage{comment,verbatim}
\usepackage{bold-extra,longtable,mdwlist}
\usepackage[colorlinks=true, allcolors=blue]{hyperref}
\usepackage[font=small, labelfont=bf, width=0.8\textwidth]{caption}
\usepackage{physics}
\usepackage{enumitem}
\usepackage{array,booktabs}
\usepackage{accents}
\usepackage{subcaption}

\usepackage{epstopdf}
\usepackage{tikz}
\usepackage{pstricks} 
\usetikzlibrary{graphs,patterns,decorations.markings,arrows,matrix}

\theoremstyle{definition}
\newtheorem{thm}{Theorem}
\newtheorem{lemma}[thm]{Lemma}
\newtheorem{remark}[thm]{Remark}

\newtheorem{prop}[thm]{Proposition}

\numberwithin{thm}{section} 

\newcommand{\Rmnum}[1]{uppercase\expandafter{\romannumeral #1}}

\def\e{{\rm e}}\def\ii{{\rm i}}
\def\dd{{\rm d}}

\DeclareMathOperator{\res}{Res}

\def\be{\begin{equation}}
\def\ee{\end{equation}}

\begin{document}

\title{On transition probabilities for the ASEP on the half-line}

\author[1]{Alexandr Garbali}
\author[1]{Cengiz Gazi}
\author[1]{Jan de Gier}
\author[2]{William Mead}
\author[1]{Michael Wheeler}

\affil[1]{School of Mathematics and Statistics, University of Melbourne, Victoria 3010, Australia}
\affil[2]{Institute for Applied Mathematics, University of Bonn, 53115 Bonn}

\date{}

\maketitle

\begin{abstract}
We study the asymmetric simple exclusion process on the half-line with particle insertion and no particle removal at the boundary. We point out that there exist two distinct multiple contour integral formulas for the transition probability between two finite particle configurations containing the same number of particles. We recall one of these formulas from the literature and prove a second, new formula by showing that it satisfies the evolution equation using a coordinate Bethe ansatz calculation and verify that the initial condition is implemented by the contour integrals. As an application we sum this formula over all final configurations and obtain an integral formula for the probability that the number of particles is conserved up to a given time. The proofs use symmetric rational functions and their properties together with contour deformations.
\end{abstract}

\noindent\textbf{E-mail:} alexandr.garbali@unimelb.edu.au, gazic@student.unimelb.edu.au, jdgier@unimelb.edu.au,\\ wmead@uni-bonn.de, wheelerm@unimelb.edu.au

\section{Introduction}
The asymmetric simple exclusion process (ASEP) is a fundamental exactly solvable interacting particle system in non-equilibrium statistical mechanics. It is closely related to the XXZ quantum spin chain and the (stochastic) six vertex model for both periodic and open boundaries \cite{GwaSpohn1992,deGierEssler2005}. The ASEP describes particles on the one dimensional lattice $\mathbb Z$ that jump to neighbouring sites subject to an exclusion rule. This model is a solvable lattice model of a Markov process that is a canonical representative of the Kardar--Parisi--Zhang (KPZ) universality class.

Exact formulas for ASEP transition probabilities go back to Sch{\"u}tz and to Tracy and Widom. Sch{\"u}tz \cite{shutz_exactsol_1997} considered the totally asymmetric simple exclusion process (TASEP), the specialisation of ASEP in which particles jump in only one direction. Using the coordinate Bethe ansatz Sch{\"u}tz solved the master equation for the $n$-particle TASEP on $\mathbb Z$ and obtained an explicit contour integral expression for the probability of observing particles at positions $(\nu_1 \ldots \nu_n)$ at time $t>0$ given that they started at positions $(\mu_1 \ldots \mu_n)$ at time $0$. Tracy and Widom extended this result to ASEP in \cite{tracy_intformula_2008}. In a subsequent work \cite{tracy_bose_2013} they generalised their integral expression to ASEP on the half-line $\mathbb Z_{>0}$ with neither particle insertion nor removal; we refer to these as closed boundary conditions. Other works on half-line ASEP include \cite{barraquand_2018,barraquand_2024,garbali_symmetric_2023,de_gier_2025}.

In this paper, we consider ASEP on the half-line with particle insertion at site $1$ but without particle removal. We propose and prove an integral formula for the transition probability when the initial and final configurations contain the same number of particles. This formula is given by a multiple integral of an expression given by a summation over signed permutations or the hyperoctahedral group, also know as the Weyl group of type $\mathrm B$. 

By summing this formula over all final configurations with the same particle number we also obtain a formula for the probability that the number of particles is conserved up to time $t$. This formula is given by a multiple integral of a completely factorised function. An important feature of our proofs is that technical residue calculations are governed by properties of the the off-shell Bethe wave function which we regards as symmetric rational function $F_\nu$, which appears in the transition probability integrand. The properties of this function such as recurrence relations, vanishing and divisibility under specializations of the variables are used to prove the initial condition and to control the contour deformations required when summing over final configurations. More generally these arguments suggest an approach to integrable stochastic particle systems for which transition probabilities admit multiple contour integral representations in terms of symmetric functions. Specifically, certain analytic questions about these integral formulas can be reduced to algebraic properties of the symmetric functions entering their integrands.

\subsection{ASEP on the half-line}
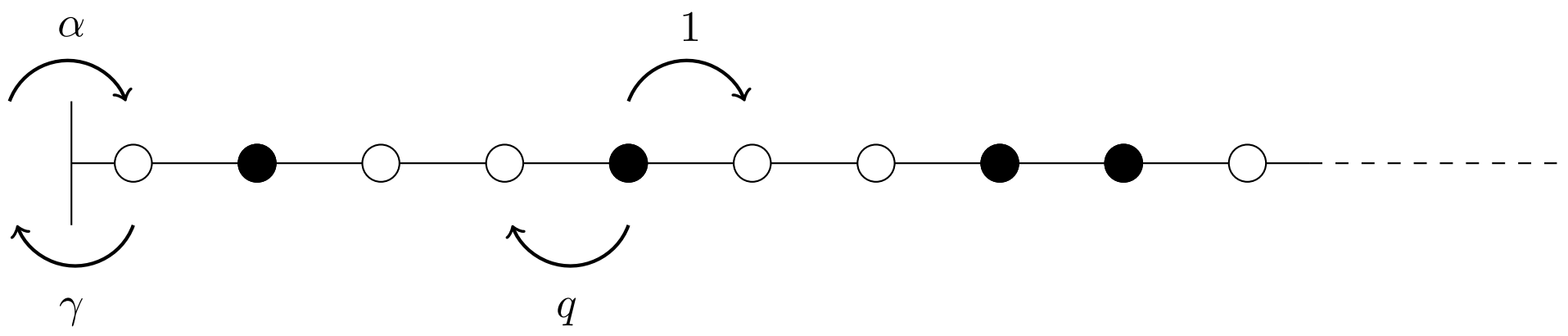
\begin{figure}[h]
\begin{center}
\begin{tikzpicture}[scale=1]
 \draw(0.5,0) -- (10.5,0);
 \draw(0.5,-0.5) -- (0.5,0.5);
 \draw[dashed] (10.5,0) -- (12.5,0);
 
  \foreach \i in {0,...,9}
    {
    \pgfmathsetmacro{\xcoord}{1+\i))}
    \pgfmathsetmacro{\ycoord}{0}
    \draw[fill,color=white,draw=black] (\xcoord,\ycoord) circle (0.15);
    }
 \draw[fill,color=black] ({1+1},{0}) circle (0.15);
\draw[fill,color=black] ({1+4},{0}) circle (0.15);
\draw[fill,color=black] ({1+7},{0}) circle (0.15);
\draw[fill,color=black] ({1+8},{0}) circle (0.15);

\draw[thick,->] (5,0.5) arc (160:20:0.5) ;
\draw[thick,->] (5,-0.5) arc (-20:-160:0.5) ;
\node[scale=1] at (5.5,1.1){$1$};
\node[scale=1] at (4.5,-1.2){$q$};

\draw[thick,->] (0,0.5) arc (160:20:0.5) ;
\draw[->,thick] (1,-0.5) arc (-20:-160:0.5);
\node[scale=1] at (0.5,1.1){$\alpha$};
\node[scale=1] at (0.5,-1.2){$\gamma$};

\end{tikzpicture}
\end{center}
\caption{Dynamics of the ASEP on the half-line $\mathbb{N}$.}
    \label{ASEP dynamics figure}
\end{figure}

We consider the continuous-time ASEP on the semi-infinite half-line $\mathbb{N}:=\mathbb{Z}_{>0}$ with configurations given by finitely many particles. Let $\mathbb{W}$ denote the set of finite subsets of $\mathbb{N}$, then the particle configurations are indexed by the random variable $\nu = (\nu_1(t),\dots,\nu_n(t)) \in \mathbb{W}$ where $\nu_1 > \nu_2 > \cdots > \nu_n$ for some finite $n \geq 0$ and the numbers $\nu_i$ are interpreted as the locations of particles on the half-line. The particle configurations in the half-line ASEP evolve according to the bulk transition rates 
\begin{align*}
	\nu \mapsto (\nu_1,\dots,\nu_i+1,\dots,\nu_n) & \text{ at rate $1$ if } \nu_{i-1}>\nu_i +1 \\
	\nu \mapsto (\nu_1,\dots,\nu_i-1,\dots,\nu_n) & \text{ at rate $q$ if } \nu_{i+1}<\nu_i -1
\end{align*}
where we put $\nu_0=\infty$ and $\nu_{n+1}=0$ and boundary transition rates 
\begin{align*}
(\nu_1,\dots,\nu_n) \mapsto (\nu_1,\dots,\nu_n,1) & \text{ at rate $\alpha$ if } \nu_{n}>1 \text{ or } n=0 
\\(\nu_1,\dots,\nu_n) \mapsto (\nu_1,\dots,\nu_{n-1}) & \text{ at rate $\gamma$ if } \nu_{n}=1
\end{align*}
The bulk and boundary dynamics are depicted in Figure \ref{ASEP dynamics figure}. In this work we restrict to the case $\gamma=0$, therefore only insertions of particles are allowed at location $1$ on the half-line.  

For fixed $\nu\in\mathbb{W}$, let $\nu^{i,i+1}\in\mathbb{W}$ be the configuration obtained from $\nu$ by interchanging the occupations of sites $i$ and $i+1$. Define $\eta_i=1$ if $i\in\nu$ and $\eta_i=0$ otherwise. The Markov generator of half-line ASEP with $\gamma=0$ is the operator $\mathscr{L}$ acting on functions $f:\mathbb{W}\to\mathbb{C}$ as follows
\begin{align}
    \label{half-line ASEP generator}
    \mathscr{L}[f](\nu) &= 
    \alpha \eta_1 
    f(\nu\setminus\{1\})
    + 
    \sum_{s=1}^\infty 
    \left((1-\eta_s) \eta_{s+1} + q (1-\eta_{s+1}) \eta_{s} \right)
    f(\nu^{s,s+1}) \nonumber \\
    &-\alpha(1- \eta_1) 
    f(\nu)
    -
    \sum_{s=1}^\infty 
    \left( q (1-\eta_s) \eta_{s+1} + (1-\eta_{s+1}) \eta_{s} \right)
    f(\nu)
\end{align}
Using the Markov generator $\mathscr{L}$ we write the evolution equation of the ASEP on the half-line for a function $\psi_t:\mathbb{W} \to \mathbb{R}$
\begin{equation}
	\label{half-line ASEP evolution equation}
	\frac{\dd}{\dd t} \psi_t(\nu) = \mathscr{L}[\psi_t](\nu)
\end{equation}

Let $\mu,\nu\in \mathbb{W}$ be arbitrary particle configurations. We will denote by $\mathbb{P}_t(\mu\to\nu)$ a solution to \eqref{half-line ASEP evolution equation} for which the particles at time $t>0$ are located at positions given by $\nu$ and subject to the initial condition in which the particles at $t=0$ are  located at the positions given by $\mu$, i.e.
\begin{align}\label{initial condition}
\mathbb{P}_0(\mu\to\nu) = \delta_{\mu,\nu}
\end{align}
We refer to $\mathbb{P}_t(\mu\to\nu)$ as the \emph{transition probability}. 

\subsection{Transition probability}
A formula for the transition probability of half-line ASEP with closed boundary conditions ($\alpha=\gamma=0$) was given in \cite{tracy_bose_2013}. Here we rederive the transition probability for open boundary conditions with $\alpha>0$ and $\gamma=0$, under the additional condition that the number of particles at time $t$ equals the number at time zero. Closed boundary conditions are recovered by setting $\alpha=0$.

Define the \emph{one-particle wave function} $\phi$ and its eigenvalue $\lambda$ by
\begin{equation}\label{eq:phi lambda}
    \phi(\nu,u) = \frac{1}{u-q^{1/2}} \left(\frac{q^{1/2} - u}{1-q^{1/2} u}\right)^{\nu-1},\qquad \lambda(u) = -\frac{(1-q)^2 }{(1-q^{1/2} u)(1-q^{1/2} u^{-1})}
\end{equation}
The \emph{two-particle scattering factor} $S$ and the \emph{boundary scattering factor} $K$ are defined by
\begin{equation}\label{eq: S K}
    S(u,v) = \frac{u-v}{u-q v},\qquad K(u) = \frac{(1- u^2)(u-q^{1/2})}{(1-q - \alpha(1-q^{-1/2} u))(1-q^{1/2} u)}\,
\end{equation}
In some formulas, it is convenient to use the parameter
\begin{equation}
    a = -\frac{\alpha}{q^{1/2}(1-q-\alpha)}
\end{equation}
in place of $\alpha$. In terms of this parameter, the boundary scattering factor $K(u)$ is
\begin{equation}\label{eq: K a}
    K(u) = \frac{(1-u^2)(u-q^{1/2})(1-aq^{1/2})}{(1-q)(1-au)(1-q^{1/2} u)}
\end{equation}

We denote the Weyl groups of types $A$ and $B$ by $W_n^\mathrm{A}$ and $W_n^\mathrm{B}$, respectively. Here \(W_n^{\mathrm A}\cong S_n\) is the symmetric group, whereas
\(W_n^{\mathrm B}\cong \{\pm 1\}^n\rtimes S_n\) is the group of signed permutations. In the following we shall use symmetrisation over these Weyl groups, for example
\begin{equation}
\sum_{\sigma\in W_n^\mathrm{B}} \sigma\cdot \left( F(u_1 \ldots u_n)\right) := \sum_{\sigma\in W_n^\mathrm{B}} F(u_{\sigma_1} \ldots u_{\sigma_n})
\end{equation}
where the sum is over signed permutations and $u_{-i} := u_i^{-1}$ for $i \in \{1 \ldots n\}$. In what follows, $\mathbb{W}_k\subset \mathbb W$ denotes the set of subsets of size $k\geq 0$.
\begin{prop}
\label{prop1}
	Let $\mu,\nu \in\mathbb{W}_n$ be two ASEP configurations with the same number of particles. Under the specialisation $\gamma=0$, the ASEP transition probability is given by
    \begin{multline}
    \label{eq:Ptnu}
		\mathbb{P}_t(\mu\to\nu)= (1-q)^n q^{1/2\sum_{i=1}^n (\mu_i-\nu_i)} \mathrm{e}^{-\alpha t} \oint \frac{\dd u_1}{2\pi\ii u_1} \cdots \oint \frac{\dd u_n}{2\pi\ii u_n} \prod_{i=1}^n \e^{\lambda(u_i)t} \\ \times  \sum_{\sigma\in W_n^\mathrm{B} } \sigma\cdot \left( \prod_{1\leq i<j\leq n}S(u_i,u_j)S(u_i,u_j^{-1}) \prod_{i=1}^n K(u_i)\phi(\nu_i,u_i) 
		\right) \\ \times \prod_{1\leq i<j\leq n}
		\frac{1}{S(u_i,u_j)S(u_i,u_j^{-1})}
        \prod_{i=1}^n \frac{-1}{K(u_i)} \phi(\mu_i,u_i^{-1}),
	\end{multline}
where the contours encircle $u_i=q^{-1/2}$ and no other poles.
\end{prop}
We prove this proposition in Section \ref{proof of transition probability}. Note that symmetrisation in \eqref{eq:Ptnu} is over the wave functions corresponding to the final state with coordinates $\nu$. Equation \eqref{eq:Ptnu} for $\alpha=0$ is very similar to the one given in \cite{tracy_bose_2013}. The following is an expression for the ASEP transition probability with $\alpha\ge 0$ and $\gamma=0$ when $m=|\mu|\le |\nu|=n$. It corresponds to the homogeneous ASEP specialization of Conjecture~5.5 in \cite{garbali_symmetric_2023} proved in \cite{de_gier_2025}.
\begin{prop}[\cite{garbali_symmetric_2023,de_gier_2025}]
Let $\mu=(\mu_1 \ldots \mu_m)$ with $\mu_i > \mu_{i+1}$ denote the positions of $m$ particles at time zero, and $\nu=(\nu_1 \ldots \nu_n)$ with $\nu_i > \nu_{i+1}$ the positions at time $t$. The ASEP transition probability under the specialisation $\gamma=0$ and $m\leq n$ is given by
\begin{multline}
\label{eq:Ptmu}
		\mathbb{P}_t(\mu\to\nu) = V_{n, m}(q) q^{1/2(n-m - \sum_{i=1}^n \nu_i+ \sum_{i=1}^m \mu_i)} \mathrm{e}^{-\alpha t} \oint \frac{\dd u_1}{2\pi\ii u_1} \cdots \oint \frac{\dd u_n}{2\pi\ii u_n} \prod_{i=1}^n \e^{\lambda(u_i)t}  \\ \times \prod_{1\leq i<j\leq n}S(u_i,u_j)S(u_i,u_j^{-1}) 
		\prod_{i=1}^n K(u_i)\phi(\nu_i,u_i)
		\\ \times \sum_{\sigma\in W_n^\mathrm{B} } \sigma\cdot \left(\prod_{1\leq i<j\leq n}
		\frac{1}{S(u_i,u_j)S(u_i,u_j^{-1}) } \prod_{i=1}^m \frac{1}{K(u_i)} \phi(\mu_i,u_i^{-1})
		\right),
\end{multline}
with contours surrounding $u_i=q^{1/2}$ and
\begin{equation}
V_{n,m}(q) = 
\left\{
\begin{array}{ll}
\displaystyle
\dfrac{\alpha^{n-m} (1-q^{1/2})^n}{2\left(1-q^{1/2(n-m)} \right)\left(q;q\right)_{n-m-1}}q^{1/2\binom{n-m}{2}}  \qquad\qquad & n>m \\[5mm]
\displaystyle (1-q^{1/2})^n & n=m
\end{array}\right.
\end{equation}
\end{prop}

\begin{remark}
    Formula \eqref{eq:Ptnu} is different in nature from the formula for $\mathbb{P}_t(\mu\to\nu)$ in \eqref{eq:Ptmu} for which the sum over $W^B_n$ occurs over the factors related to the initial coordinates $\mu$ instead of over those related to $\nu$ as in \eqref{eq:Ptnu} . In the case $|\mu|=|\nu|$ the formulas give the same result after computing residues.
\end{remark}

\subsection{The particle number conservation probability}
Let us consider the following application of the Proposition \ref{prop1}. At time $0$ we start with the system in an arbitrary $n$-particle configuration $\mu=(\mu_1 \ldots \mu_n)$ and ask what is the probability that the system retains an $n$-particle configuration at time $t>0$. We denote this quantity by $C_t(\mu)$. It can be computed by summing $\mathbb{P}_t(\mu\to\nu)$ over all possible $n$-particle states $\nu=(\nu_1 \ldots \nu_n)$
\begin{align}
\label{defn:C}
C_t(\mu)= \sum_{\nu\in \mathbb W_n}\mathbb{P}_t(\mu\to \nu)
\end{align}
The summation can be computed using expression \eqref{eq:Ptnu}.
\begin{prop}\label{prop:Pt_nparticles} Consider the half-line ASEP under the specialisation $\gamma = 0$ and with the initial configuration of particles at $t=0$ given by $\mu=(\mu_1 \ldots \mu_n)$. The probability $C_t(\mu)$ that at time $t>0$ there are $n$ particles in the system is given by
\begin{multline}\label{integral formula for C}
C_t(\mu)= \alpha^n q^{1/2(\sum_{i=1}^n \mu_i-n)} \mathrm{e}^{-\alpha t} \oint \frac{\dd u_1}{2\pi\ii u_1} \cdots \oint \frac{\dd u_n}{2\pi\ii u_n} \prod_{i=1}^n \e^{\lambda(u_i)t} \\ \times\prod_{1\leq i<j\leq n}
		S(u_j,u_i)S(u_j^{-1},u_i)
        \prod_{i=1}^n K(u_i^{-1}) \phi(\mu_i,u_i^{-1}),
\end{multline}
where the contours encircle $u_i=q^{-1/2}$ and $u_i=0$ and no other poles.
\end{prop}

\begin{remark}
The proof of \eqref{integral formula for C} uses a symmetrisation identity that combines the summation over the final coordinates $\nu$ in \eqref{defn:C} with the sum over the Weyl group $W_n^\mathrm{B}$ in \eqref{eq:Ptnu}. We stress that such a symmetrisation identity is not available if we choose to use expression \eqref{eq:Ptmu}. The remaining challenge is to interchange the summation over $\nu$ with the contour integrals in \eqref{eq:Ptnu}, which is nontrivial but can be achieved. 
\end{remark}

\subsection{Outline}
Section \ref{section: F} is devoted to the discussion of a symmetric function $F_\nu(u)$ and its properties. Many technical aspects in our proofs will rely on the results of this section. In Section  \ref{proof of transition probability} we prove Proposition \ref{prop1} by showing that our integral formula for $\mathbb{P}_t(\mu\to\nu)$ satisfies the evolution equation and the initial condition. In Section \ref{proof of n-particle probability} we prove Proposition \ref{prop:Pt_nparticles} by a direct computation starting from our integral expression for $\mathbb{P}_t(\mu\to\nu)$.

\subsection{Acknowledgments}
We gratefully acknowledge support from the Australian Research Council. WM was supported by the Deutsche Forschungsgemeinschaft (DFG, German Research Foundation) under Germany's Excellence Strategy - GZ 2047/1, Projekt-ID 390685813. CG is supported by the Fay Marles scholarship. AI was used for minor proofreading and references.


\section{Symmetric function $F_\nu(u_1\ldots u_n)$}\label{section: F}
In this section, we discuss a $W_n^A$-symmetric rational function $F_\nu(u)$ that plays a central role in our proofs. This function is the off-shell coordinate Bethe ansatz wave function for the XXZ spin chain \cite{Lieb1967,baxterbook}. It can be written as a partition function of the six-vertex model using the language of algebraic Bethe ansatz \cite{FaddeevTakhtajan1979}, see also \cite{Izergin1987,Korepin1982,KitanineMailletTerras1999}. It has received recent attention in the context of symmetric functions, see for example \cite{borodin_petrov_2018}, and is the homogeneous limit of the symmetric rational function defined by this partition function. Many results are known about this function, however, we make this section self-contained and give proofs of all statements which we require in the following sections. We give a recurrence relation for $F_\nu(u)$, establish the properties of $F_\nu(u)$ that arise from specialisations of its arguments and derive a Littlewood-type summation identity. These results will be used in the subsequent sections. 

For $\nu \in\mathbb W_n$ we define a rational function
\begin{align}
    \label{eq:F}
    F_\nu(u) := F_\nu(u_1\ldots u_n) = 
    \sum_{\sigma\in W_n^\mathrm{A} }  \prod_{1\leq i<j\leq n}S(u_{\sigma(i)},u_{\sigma(j)})\prod_{i=1}^n \phi(\nu_i,u_{\sigma(i)}).
\end{align}
This function admits a recursive formula which also shows its explicit dependence on a particular argument, say $u_n$. This formula is obtained from \eqref{eq:F} where we write the sum $\sigma\in W_n^\mathrm{A}$ by fixing $\sigma(k)=n$ and summing over $k=1\ldots n$
\begin{multline}
    \label{eq:F sym fix n 0}
    F_\nu(u)=
    \sum_{k=1}^n
    \phi(\nu_k,u_{n})
    \sum_{\substack{\sigma\in W_n^\mathrm{A}\\\sigma(k)=n}}
         \prod_{j=k+1}^nS(u_{n},u_{\sigma(j)})
     \\ \times \prod_{i=1}^{k-1}S(u_{\sigma(i)},u_{n})
     \prod_{\substack{1\leq i<j\leq n\\i,j\neq k}}S(u_{\sigma(i)},u_{\sigma(j)})\prod_{\substack{i=1\\i\neq k}}^n \phi(\nu_i,u_{\sigma(i)}).
\end{multline}
Next we note that the summation over $\sigma\in W_n^\mathrm{A}$ with fixed $\sigma(k)=n$ can be written using a summation over subsets $A\subseteq [1\ldots n-1]$ of fixed lengths $|A|=k-1$ and summations over permutations of $A$ and $A^c$ (where $A^c$ is the complement of $A$ w.r.t. $[1\ldots n-1]$). The latter summations give rise to two functions $F$ and we obtain
\begin{multline}\label{eq:F rec}
     F_\nu(u)=
    \sum_{k=1}^n
    \phi(\nu_k,u_{n})
    \sum_{\substack{A\subseteq [1\ldots  n-1]\\|A|=k-1}}
         \prod_{i\in A^c}S(u_{n},u_{i})
     \prod_{i\in A}S(u_{i},u_{n})
     \prod_{\substack{i\in A\\j\in A^c}}S(u_i,u_j)
     \\ \times F_{\nu_1\ldots \nu_{k-1}}(u_A)F_{\nu_{k+1}\ldots \nu_{n}}(u_{A^c}),
\end{multline}
where $u_A$ is the set of arguments $u_i$ with $i\in A$ and similarly for $u_{A^c}$.

From the definition of $F_\nu(u)$ we see that it can be written as 
\begin{align}
    \label{eq:F factors}
    F_\nu(u) = \frac{\prod_{1\leq i\neq j\leq n}S(u_i,u_j) }{\prod_{i=1}^n(q^{1/2}-u_i)(1-q^{1/2} u_i)^{\nu_1-1}} \widetilde F_\nu(u),
\end{align}
where $\widetilde F_\nu(u)$ is a symmetric polynomial in $u_i$ with the top degree given by $u_i^{\nu_1-1}$. This expression tells us that $F_\nu(u)$ has a simple pole at  $u_i=q^{1/2}$, a pole of order $\nu_1-1$ at $u_i=q^{-1/2}$ and for every pair $i,j$ it has a simple pole at $u_i=q u_j$ and a double zero at $u_i=u_j$.

\begin{prop}\label{prop: wheel 1}
The residue of $F_\nu(u)$ at $u_k=q u_l$ is regular at $u_l=q^{-1/2}$. 
\end{prop}
\begin{proof}
To see this we compute this residue using \eqref{eq:F} and assume $k<l$ (the case $k>l$ follows from the symmetry of $F_\nu(u)$ in the arguments $u_k$ and $u_l$)
\begin{multline}
    \label{eq:F res}
     \underset{u_k=q u_l}{\res} F_\nu(u) = 
     \sum_{\substack{\sigma\in W_n^\mathrm{A}\\ \sigma^{-1}(k)<\sigma^{-1}(l)}}  
     \underset{u_k=q u_l}{\res} 
     \left(
     \prod_{1\leq i<j\leq n}S(u_{\sigma(i)},u_{\sigma(j)})\right) 
     \\ \times \prod_{\substack{i=1\\ i\neq k,l}}^n \phi(\nu_{\sigma^{-1}(i)},u_{i})
    \cdot     \phi(\nu_{\sigma^{-1}(k)},q u_{l})
     \phi(\nu_{\sigma^{-1}(l)},u_{l}),
\end{multline}
where the condition $\sigma^{-1}(k)<\sigma^{-1}(l)$ keeps only non-vanishing terms. Next we evaluate the product of the last two factors using the explicit form of $\phi$ from \eqref{eq:phi lambda}
\begin{align*}
    \phi(\nu_{\sigma^{-1}(k)},q u_{l})
     \phi(\nu_{\sigma^{-1}(l)},u_{l}) = 
     (1-q^{1/2}u_l)^{\nu_{\sigma^{-1}(k)}-\nu_{\sigma^{-1}(l)}-1}\frac{(q^{1/2}-u_l)^{\nu_{\sigma^{-1}(l)}-2}}{(1-q^{3/2}u_l)^{\nu_{\sigma^{-1}(k)}-1}}.
\end{align*}
The value $\nu_{\sigma^{-1}(k)}-\nu_{\sigma^{-1}(l)}-1$ is non-negative because the summation in \eqref{eq:F res} runs over permutations with the property $\sigma^{-1}(k)<\sigma^{-1}(l)$ and the particle ordering is such that $\nu_a>\nu_b$ for $a<b$. Therefore \eqref{eq:F res} is regular at $u_l=q^{-1/2}$. 
\end{proof}

\begin{prop}\label{prop: wheel 2}
Let $n\geq 2$ and $k=1\ldots n$, we have the property
\begin{align}
    \label{eq:wheel conditions of F}
    \widetilde F_\nu(\ldots u_k=q^{\pm 1/2}\ldots ) = \prod_{\substack{j=1\\ j\neq k}}^n (1-q^{\pm 1/2} u_j) \, \widetilde F^\pm_\nu(\ldots \hat{u}_k \ldots )
\end{align}
where $\widetilde F^\pm_\nu(\ldots \hat{u}_k \ldots )$ is a polynomial in $u_1\ldots u_{k-1},u_{k+1}\ldots u_n$. 
\end{prop}
\begin{proof}
We write $\widetilde F_\nu(u)$ in terms of $F_\nu(u)$ using \eqref{eq:F factors} and then substitute \eqref{eq:F} and the explicit formula for $\phi(\nu,u)$
\begin{multline}
\label{eq:F tilde sym}
    \widetilde F_\nu(u) =
     \frac{\prod_{i=1}^n(q^{1/2}-u_i)(1-q^{1/2} u_i)^{\nu_1-1}} {\prod_{1\leq i\neq j\leq n}S(u_i,u_j) } \\ \times
    \sum_{\sigma\in W_n^\mathrm{A} }  \prod_{1\leq i<j\leq n}S(u_{\sigma(i)},u_{\sigma(j)})\prod_{i=1}^n
    \frac{1}{u_{\sigma(i)}-q^{1/2}} \left(\frac{q^{1/2} - u_{\sigma(i)}}{1-q^{1/2} u_{\sigma(i)}}\right)^{\nu_i-1}.
\end{multline}
If $\nu_n>1$ (and hence all $\nu_i>1$), then every term in the sum over $\sigma$ in \eqref{eq:F tilde sym} is divisible by $(q^{1/2}-u_i)^{\nu_n-2}$. By combining this with the factor in front of the summation over $\sigma$ in \eqref{eq:F tilde sym} we find that the polynomial $\widetilde F_\nu(u)$ is divisible by
\begin{align}\label{eq:wheel ++}
\prod_{i=1}^n (q^{1/2}-u_i)^{\nu_n-1}.
\end{align}
Therefore \eqref{eq:wheel conditions of F} is satisfied for $\nu_n>1$ for both specializations $u_k=q^{\pm 1/2}$. Next we consider the case $\nu_n=1$. Let us fix $k=n$ for convenience (the case $k\neq n$ follows from the symmetry of $\widetilde F_\nu(u)$ in $u_i$). From \eqref{eq:F tilde sym} it can be seen that after setting $u_n=q^{1/2}$ all terms with $\sigma(n)\neq n$ will vanish while the terms with $\sigma(n)=n$ will remain finite thanks to the factor $(q^{1/2}-u_n)$ in front of the summation over $\sigma$.
This gives us
\begin{multline}
    \label{eq:F tilde sym spec 1}
    \widetilde F_{(\nu_1\ldots \nu_{n-1},1)}(u_1\ldots u_{n-1},q^{1/2})=
    \frac{-(1-q)^{\nu_1-1}} {\prod_{i=1}^{n-1} S(q^{1/2},u_i) }
    \frac{\prod_{i=1}^{n-1}(q^{1/2}-u_i)(1-q^{1/2} u_i)^{\nu_1-1}} {\prod_{1\leq i\neq j\leq n-1}S(u_i,u_j) }
    \\
    \sum_{\sigma\in W_{n-1}^\mathrm{A} }  \prod_{1\leq i<j\leq n-1}S(u_{\sigma(i)},u_{\sigma(j)})\prod_{i=1}^{n-1}
    \frac{1}{u_{\sigma(i)}-q^{1/2}}
     \left(\frac{q^{1/2} - u_{\sigma(i)}}{1-q^{1/2}u_{\sigma(i)}}\right)^{\nu_i-1}.
\end{multline}
By comparing this with \eqref{eq:F tilde sym} we see that the right hand side can be written in terms of $\widetilde F_{(\nu_1\ldots \nu_{n-1})}(u_1\ldots u_{n-1})$
\begin{align}
    \label{eq:Ft rec}
    \widetilde F_{(\nu_1\ldots \nu_{n-1},1)}(u_1\ldots u_{n-1},q^{1/2})=
    \frac{-(1-q)^{\nu_1-1}} {\prod_{i=1}^{n-1} S(q^{1/2},u_i) }
    \widetilde F_{(\nu_1\ldots \nu_{n-1})}(u_1\ldots u_{n-1}).
\end{align}
Since $\widetilde F_\nu(u)$ is a polynomial and 
\begin{align*}
    \frac{1} {\prod_{i=1}^{n-1} S(q^{1/2},u_i) } = \prod_{i=1}^{n-1}q^{1/2} \frac{1-q^{1/2} u_i}{q^{1/2} - u_i},
\end{align*}
it follows that $\widetilde F_{(\nu_1\ldots \nu_{n-1},1)}(u_1\ldots u_{n-1},q^{1/2})\propto  \prod_{i=1}^{n-1}(1-q^{1/2} u_i)$. By symmetry this implies \eqref{eq:wheel conditions of F}  for the case $u_k=q^{1/2}$. Because of this we have a vanishing condition
$$
    \widetilde F_\nu(\ldots u_k=q^{1/2}\ldots u_l=q^{-1/2} \ldots) = 0.
$$
Since $\widetilde F_\nu$ is a polynomial this means that $\widetilde F_\nu(\ldots u_l=q^{-1/2} \ldots)$ has a zero at $u_k=q^{1/2}$ and thus divisible by $u_k-q^{1/2}$ for any $k\neq l$. This implies \eqref{eq:wheel conditions of F}  for the case $u_k=q^{-1/2}$ and completes the proof.
\end{proof}

\begin{prop}\label{prop: sum nu}
    Consider $n$ parameters $u_i\in \mathbb C$, $i=1\ldots n$, such that
    \begin{align}\label{eq:convergence condition}
        \left|\frac{1 -q^{-1/2} u_i}{1-q^{1/2} u_i} \right|<1,\qquad i=1\ldots n,
    \end{align}
    then we have the following identity
    \begin{align}
        \label{eq:F summed}
      \sum_{\nu\in \mathbb W_n}
q^{-1/2\sum_{i=1}^n \nu_i} F_\nu(u_1\ldots u_n)=  
\frac{(-1)^n}{(1-q)^n}\prod_{i=1}^n \frac{(1-q^{1/2} u_i)}{(q^{1/2}- u_i)u_i}\prod_{1\leq i\neq j\leq n}S(u_i,u_j).
    \end{align}
\end{prop}
\begin{proof}
We use the definition of $F_\nu(u)$ \eqref{eq:F} and write the sum
\begin{align}\label{eq:sum F proof}
\sum_{\nu\in \mathbb W_n}
q^{-1/2\sum_{i=1}^n \nu_i} F_\nu(u_1\ldots u_n) =
\sum_{\sigma\in W_n^\mathrm{A} }  \prod_{1\leq i<j\leq n}S(u_{\sigma(i)},u_{\sigma(j)})
\sum_{\nu\in \mathbb W_n}
q^{-1/2\sum_{i=1}^n \nu_i} \prod_{i=1}^n \phi(\nu_i,u_{\sigma(i)}).
\end{align}
We recall the definition of the function $\phi$ from \eqref{eq:phi lambda} and note that the summation on the right hand side can be computed by recursively summing a geometric series. We may use the formula 
\begin{align}\label{eq:nested geometric series}
    \sum_{\nu\in \mathbb W_n}
\prod_{i=1}^n x_i^{-\nu_i} = \prod_{i=1}^n\frac{-1}{1-\prod_{j=1}^i x_j}, 
\qquad |x_1^{-1}\cdots x_j^{-1}|<1~~\text{for } j=1\ldots n,
\end{align}
and compute the sum
\begin{align}
    \sum_{\nu\in \mathbb W_n}
q^{-1/2\sum_{i=1}^n \nu_i} \prod_{i=1}^n \phi(\nu_i,u_{i})&=
    \sum_{\nu\in \mathbb W_n}
q^{-1/2\sum_{i=1}^n \nu_i} \prod_{i=1}^n  \frac{1}{u_i-q^{1/2}} \left(\frac{q^{1/2} - u_i}{1-q^{1/2} u_i}\right)^{\nu_i-1}
\nonumber\\
&=(-1)^n\prod_{i=1}^n  \frac{1-q^{1/2} u_i}{(q^{1/2} - u_i)^2}
\times\sum_{\nu\in \mathbb W_n}
 \prod_{i=1}^n   \left(\frac{1-q^{1/2} u_i}{1 -q^{-1/2} u_i}\right)^{-\nu_i}
 \nonumber\\
&=(-1)^n\prod_{i=1}^n  \frac{1-q^{1/2} u_i}{(q^{1/2} - u_i)^2}
\times \prod_{i=1}^n\frac{-1}{1-\prod_{j=1}^i \frac{1 -q^{1/2} u_j}{1-q^{-1/2} u_j}},
\label{eq:nu sum computed}
\end{align}
where we used \eqref{eq:nested geometric series} to get the expression in the last line. For the convergence we require the condition \eqref{eq:convergence condition}. This is a stronger condition than the condition in \eqref{eq:nested geometric series} but it is necessary for the convergence of every term in the summation over permutations $\sigma$ in \eqref{eq:sum F proof}. Using \eqref{eq:nu sum computed} in \eqref{eq:sum F proof} gives us
\begin{multline}\label{eq:summed F proof}
\sum_{\nu\in \mathbb W_n}
q^{-1/2\sum_{i=1}^n \nu_i} F_\nu(u_1\ldots u_n) \\ =\prod_{i=1}^n  \frac{1-q^{1/2} u_i}{(q^{1/2} - u_i)^2}
\sum_{\sigma\in W_n^\mathrm{A} }  \prod_{1\leq i<j\leq n}S(u_{\sigma(i)},u_{\sigma(j)})
 \prod_{i=1}^n\frac{1}{1-\prod_{j=1}^i \frac{1 -q^{1/2} u_{\sigma(j)}}{1-q^{-1/2} u_{\sigma(j)}}}.
\end{multline}

It remains to factorise the sum over $\sigma$ on the right-hand side of
\eqref{eq:summed F proof}. Set
\[
    H_n(u_1 \ldots u_n)=
    \sum_{\sigma\in W_n^{\mathrm A}}
    \prod_{1\leq i<j\leq n}S(u_{\sigma(i)},u_{\sigma(j)})
 \prod_{i=1}^n\frac{1}{1-\prod_{j=1}^i \frac{1 -q^{1/2} u_{\sigma(j)}}{1-q^{-1/2} u_{\sigma(j)}}}
\]
We claim that
\begin{equation}
    \label{eq:permutation sum factorisation}
    H_n(u_1 \ldots u_n)
    =\frac{(-1)^n}{(1-q)^n}
      \prod_{i=1}^n\frac{q^{1/2}-u_i}{u_i}
      \prod_{1\leq i\neq j\leq n}S(u_i,u_j)
\end{equation}
We prove this identity by induction on $n$. The case $n=0$ is immediate.
Grouping the permutations according to their last entry gives
\begin{equation}
    \label{eq:H recursion}
    H_n(u_1 \ldots u_n)
    =\frac{1}{1-\prod_{i=1}^n \frac{1 -q^{1/2} u_i}{1-q^{-1/2} u_i}}
      \sum_{k=1}^n
      \left(\prod_{i\neq k}S(u_i,u_k)\right)
      H_{n-1}(u_1 \ldots \widehat{u}_k \ldots u_n)
\end{equation}
where the hat indicates omission. Substitution of the induction hypothesis
into \eqref{eq:H recursion} reduces proving \eqref{eq:permutation sum factorisation}
to proving the identity
\begin{equation}
    \label{eq:partial fraction identity for H}
    \sum_{k=1}^n
      \frac{u_k}{q^{1/2}-u_k}
      \prod_{i\neq k}\frac{u_k-q u_i}{u_k-u_i}
    =-\frac{1}{1-q}\left(1-\prod_{i=1}^n \frac{1 -q^{1/2} u_i}{1-q^{-1/2} u_i}\right).
\end{equation}
The product appearing on the right hand side is a specialization $z=q^{1/2}$ of the function
\[
    R(z)=\prod_{i=1}^n\frac{z-q u_i}{z-u_i}
\]
This function has a partial fraction decomposition
\[
    R(z)=\prod_{i=1}^n\frac{z-q u_i}{z-u_i}=1+(1-q)\sum_{k=1}^n
      \frac{u_k}{z-u_k}
      \prod_{i\neq k}\frac{u_k-q u_i}{u_k-u_i}
\]
which, after setting $z=q^{1/2}$, proves \eqref{eq:partial fraction identity for H}. Finally, substituting \eqref{eq:permutation sum factorisation} into
\eqref{eq:summed F proof} gives
\[
\begin{split}
\sum_{\nu\in\mathbb W_n}q^{-\frac12\sum_i\nu_i}F_\nu(u_1 \ldots u_n)
&=\prod_{i=1}^n\frac{1-q^{1/2} u_i}{(q^{1/2}-u_i)^2}\,
  \frac{(-1)^n}{(1-q)^n}
  \prod_{i=1}^n\frac{q^{1/2}-u_i}{u_i}
  \prod_{i\neq j}S(u_i,u_j)\\
&=\frac{(-1)^n}{(1-q)^n}
  \prod_{i=1}^n\frac{1-q^{1/2} u_i}{(q^{1/2}-u_i)u_i}
  \prod_{i\neq j}S(u_i,u_j),
\end{split}
\]
which is precisely \eqref{eq:F summed}.
\end{proof}


\section{Proof of Proposition \ref{prop1}: Transition probability for fixed number of particles}\label{proof of transition probability}
In this section we show that the expression \eqref{eq:Ptnu} satisfies the evolution equation \eqref{half-line ASEP evolution equation} and the initial condition \eqref{initial condition} with $|\mu|=|\nu|=n$. For the proof of the evolution equation \eqref{half-line ASEP evolution equation} we follow a similar coordinate Bethe ansatz approach as in \cite{Alcaraz_1987}. It is convenient to introduce several functions. Define
    \begin{multline}
    \label{eq:CF}
		\mathcal F_\nu(u;t)=\mathcal F_{\nu_1\ldots \nu_n}(u_1\ldots u_n;t) :=q^{-1/2\sum_{i=1}^n \nu_i} \e^{t\left(\sum_{i=1}^n\lambda(u_i)-\alpha\right)} \\ \times \sum_{\sigma\in W_n^\mathrm{B} } \sigma\cdot\left( \prod_{1\leq i<j\leq n}S(u_i,u_j)S(u_i,u_j^{-1}) \prod_{i=1}^n K(u_i)\phi(\nu_i,u_i) \right).
	\end{multline}
With this definition we can write
\begin{multline}
    \label{eq:Pt F}
    \mathbb{P}_t(\mu\to\nu)\big|_{|\mu|=|\nu|} = (1-q)^n q^{1/2\sum_{i=1}^n \mu_i}\oint \frac{\dd u_1}{2\pi\ii u_1} \cdots \oint \frac{\dd u_n}{2\pi\ii u_n}\mathcal F_\nu(u;t)\\ \times  \prod_{1\leq i<j\leq n}
		\frac{1}{S(u_i,u_j)S(u_i,u_j^{-1})}
        \prod_{i=1}^n \frac{-1}{K(u_i)} \phi(\mu_i,u_i^{-1}),
\end{multline}
In addition to this we note that the dependence on $\nu$ in $\mathcal F_\nu(u;t)$ is given by the function 
\begin{align}
    \label{eq:Phi}
\Phi_{\nu_1\ldots \nu_n} (u_1\ldots u_n) = \Phi_\nu (u) := q^{-1/2\sum_{i=1}^n \nu_i}  \prod_{i=1}^n \phi (\nu_i,u_i)
\end{align}
and $\mathcal F_\nu(u;t)$ is linear in $\sigma\cdot \Phi_\nu(u)$
    \begin{align}
    \label{eq:CF-Phi}
		\mathcal F_\nu(u;t)=\e^{t\left(\sum_{i=1}^n\lambda(u_i)-\alpha\right)}  \sum_{\sigma\in W_n^\mathrm{B} } \sigma\cdot\left( \Phi_{\nu}(u) \prod_{1\leq i<j\leq n}S(u_i,u_j)S(u_i,u_j^{-1}) \prod_{i=1}^n K(u_i) \right).
	\end{align}
Finally we define the function $\mathcal F_\nu(u):=\mathcal F_\nu(u;t=0)$:
    \begin{multline}
    \label{eq:CFnu}
		\mathcal F_\nu(u)=\mathcal F_{\nu_1\ldots \nu_n}(u_1\ldots u_n) := \\ q^{-1/2\sum_{i=1}^n \nu_i} \sum_{\sigma\in W_n^\mathrm{B} } \sigma\cdot\left( \prod_{1\leq i<j\leq n}S(u_i,u_j)S(u_i,u_j^{-1}) \prod_{i=1}^n K(u_i)\phi(\nu_i,u_i) \right).
	\end{multline}

\subsection{Evolution equation}
We can rewrite the evolution equation \eqref{half-line ASEP evolution equation}
in the form where the summation over the sites $s=0,1\ldots$ is replaced with the summation over the particle locations. This can be done by considering separately configurations with $\nu_n>1$ and $\nu_n=1$. 
\begin{itemize}
    \item Let $\nu=(\nu_1\ldots \nu_n)\in \mathbb W$ such that $\nu_n>1$ then \eqref{half-line ASEP evolution equation} can be written as
\begin{align}
    \label{eq: Pt_evo_bulk}
    \frac{d}{dt}\mathbb{P}_t(\nu) &= \sum_{i=1}^n \left(\mathbb{P}_t(\ldots \nu_i-1 \ldots) + 
    q \mathbb{P}_t(\ldots \nu_i+1\ldots)
    -(1+q) \mathbb{P}_t(\nu)\right)-\alpha  
    \mathbb{P}_t(\nu)
\end{align}
subject to the exclusion constraint which occurs when any pair of particles  are next to each other $\nu_l=\nu_{l+1}+1=a>1$ for $l=1\ldots n-1$
\begin{align}
    \label{eq: Pt exclusion}
    \mathbb{P}_t(\ldots a-1,a-1\ldots)+ 
    q \mathbb{P}_t(\ldots a,a\ldots) -
    (1+q) \mathbb{P}_t(\ldots a,a-1\ldots) =0
\end{align}
\item Let $\nu=(\nu_1\ldots \nu_n)\in \mathbb W$ such that $\nu_n=1$, we have
\begin{multline}
    \label{eq: Pt_evo_boundary}
    \frac{d}{dt}\mathbb{P}_t(\nu) = \sum_{i=1}^{n-1} \left(\mathbb{P}_t(\ldots \nu_i-1 \ldots) + 
    q \mathbb{P}_t(\ldots \nu_i+1\ldots)
    -(1+q) \mathbb{P}_t(\nu)\right)\\
    +q \mathbb{P}_t(\ldots \nu_{n-1},2)
     - \mathbb{P}_t(\nu) 
\end{multline}
where we set the term $\alpha \mathbb{P}_t(\ldots \nu_{n-1})$ to zero since at time $t=0$ we require to have $n$ particles and the process of removing a particle at $t>0$ at the boundary is forbidden.
\end{itemize}

\subsubsection{Proving \eqref{eq: Pt_evo_bulk}}
We prove \eqref{eq: Pt_evo_bulk} by showing that
\begin{align}
    \label{eq: CF bulk equation}
    \frac{d}{dt}\mathcal F_\nu(u;t) &= \sum_{i=1}^n \left(\mathcal F_{\ldots \nu_i-1 \ldots}(u;t) + 
    q \mathcal F_{\ldots \nu_i+1 \ldots}(u;t) 
    -(1+q) \mathcal F_{\nu}(u;t)\right) - \alpha \mathcal F_{\nu}(u;t).
\end{align}
The left hand side of this equation reads
\begin{align}\label{eq:dt CF}
    \frac{d}{dt}\mathcal F_\nu(u;t)  =\left( \sum_{i=1}^n \lambda(u_i) - \alpha\right)\mathcal F_\nu(u;t).
\end{align}
Using the explicit form of $\phi(k,u)$ given in \eqref{eq:phi lambda} we find that the function $\Phi_\nu(u)$ has the property
\begin{align}
    \label{eq:Phi shift}
\Phi_{\ldots \nu_i\pm 1\ldots}(u) = q^{\mp 1/2}
 \left(\frac{q^{1/2} - u_i}{1-q^{1/2} u_i}\right)^{\pm 1}
 \Phi_\nu(u).
\end{align}
This property gives us the identity 
\begin{multline}
    \label{eq: Phi difference}
\Phi_{\ldots \nu_i-1 \ldots}(u) + 
    q \Phi_{\ldots \nu_i+1 \ldots}(u)
    -(1+q) \Phi_{\ldots \nu_i \ldots}(u)
    \\ =
    \left(
     q^{1/2}
 \left(\frac{q^{1/2} - u_i}{1-q^{1/2} u_i}\right)^{-1}
 + q^{1/2}
 \left(\frac{q^{1/2} - u_i}{1-q^{1/2} u_i}\right)
 -(1+q)\right)\Phi_\nu (u) = \lambda(u_i)\Phi_\nu (u).
\end{multline}
where we noted that the coefficient in front of  $\Phi_\nu(u)$ in the middle expression is equal to $\lambda(u_i)$. Using \eqref{eq: Phi difference}, \eqref{eq:CF-Phi} and the fact that $\lambda(u_i)$ is invariant under the inversion of $u_i$  we have
\begin{align}
    \label{eq: CF bulk equation rhs}
    \sum_{i=1}^n \left(\mathcal F_{\ldots \nu_i-1 \ldots}(u;t) + 
    q \mathcal F_{\ldots \nu_i+1 \ldots}(u;t) 
    -(1+q) \mathcal F_{\nu}(u;t)\right) = \sum_{i=1}^n \lambda(u_i) \mathcal F_{\nu}(u;t).
\end{align}
This equation together with \eqref{eq:dt CF} proves \eqref{eq: CF bulk equation}.

\subsubsection{Proving \eqref{eq: Pt exclusion}}
Fix $l\in\{1\ldots n\}$ and $\nu_{l+1}=\nu_l=a>1$. We will prove \eqref{eq: Pt exclusion} by showing that
\begin{align}\label{eq:CF exclusion}
    \mathcal F_{\ldots a-1,a-1\ldots}+ 
    q \mathcal F_{\ldots a,a\ldots} -
    (1+q)\mathcal F_{\ldots a,a-1\ldots}=0.
\end{align}
Using \eqref{eq:Phi shift} and the definition of $S(u,v)$, given in \eqref{eq: S K}, we obtain the following expression
\begin{multline}\label{eq:Phi exclusion}
    \left(\Phi_{\ldots a-1,a-1\ldots}(u)+ 
    q \Phi_{\ldots a,a\ldots}(u) -
    (1+q)\Phi_{\ldots a,a-1\ldots}(u)\right)S(u_l,u_{l+1})
    \\ =
    \frac{q^{1/2}(1-q)(u_{l}-u_{l+1})}{(q^{1/2}-u_l)(q^{1/2}-u_{l+1})}\Phi_{\ldots a,a \ldots}(u).
\end{multline}
Write the left hand side of \eqref{eq:CF exclusion} using \eqref{eq:CF-Phi} and \eqref{eq:Phi exclusion}
\begin{multline*}
    \mathcal F_{\ldots a-1,a-1\ldots}+ 
    q \mathcal F_{\ldots a,a\ldots} -
    (1+q)\mathcal F_{\ldots a,a-1\ldots}
    =
    \e^{t\left(\sum_{i=1}^n\lambda(u_i)-\alpha\right)}\\ \times \sum_{\sigma\in W_n^\mathrm{B} } \sigma\cdot\left(
    \frac{q^{1/2}(1-q)(u_{l}- u_{l+1})}{(q^{1/2}-u_l)(q^{1/2}-u_{l+1})}\Phi_{\ldots a,a \ldots}(u)
    \frac{\prod_{1\leq i<j\leq n}S(u_i,u_j)S(u_i,u_j^{-1})}{S(u_{l},u_{l+1})} \prod_{i=1}^n K(u_i) \right).
\end{multline*}
On the right hand side of this equation $\sigma$ acts on a rational function which is manifestly antisymmetric in $u_l,u_{l+1}$, however the sum over $\sigma$ must be symmetric in $u_l,u_{l+1}$, hence it must vanish.

\subsubsection{Proving \eqref{eq: Pt_evo_boundary}}
We prove \eqref{eq: Pt_evo_boundary} by showing that $\mathcal F_\nu(u;t)$ satisfies \eqref{eq: Pt_evo_boundary}. Writing this equation in the form
\begin{align}
    \label{eq: Ft boundary}
    \frac{d}{dt}\mathcal F_\nu(u;t) = \sum_{i=1}^{n} \left(\mathcal F_{\ldots \nu_i-1 \ldots}(u;t) + 
    q \mathcal F_{\ldots \nu_i+1\ldots}(u;t)
    -(1+q) \mathcal F_\nu(u;t)\right)
     -\mathcal F_{\ldots \nu_{n-1},0}(u;t)
    +q \mathcal F_\nu(u;t),
\end{align}
and subtracting from it \eqref{eq: CF bulk equation} we get
\begin{align}
    \label{eq: F boundary equation}
    - \mathcal F_{\ldots \nu_{n-1},0}(u;t)
    + q \mathcal F_\nu(u;t)+ \alpha \mathcal F_{\nu}(u;t)=0.
\end{align}
We consider the following expression
\begin{align}\label{eq:Phi boundary}
    \left(-\Phi_{\ldots \nu_{n-1},0}(u)+ 
    q \Phi_{\ldots \nu_{n-1},1}(u)+\alpha \Phi_{\ldots \nu_{n-1},1}(u)\right)K(u_n)
    =
    \frac{q^{1/2}(1-u_n^2)}{1-q^{1/2} u_n}
    \Phi_{\ldots \nu_{n-1},1}(u).
\end{align}
The function on the right hand side is antisymmetric under the inversion of $u_n$. Combining \eqref{eq:Phi boundary} with \eqref{eq:CF-Phi} implies
\begin{multline*}
    - \mathcal F_{\ldots \nu_{n-1},0}(u;t)
    + q \mathcal F_\nu(u;t)+ \alpha \mathcal F_{\nu}(u;t)=
    \\
    \e^{t\left(\sum_{i=1}^n\lambda(u_i)-\alpha\right)}  \sum_{\sigma\in W_n^\mathrm{B} } \sigma\cdot\left(
    \frac{q^{1/2}(1-u_n^2)}{1-q^{1/2} u_n}
    \Phi_{\ldots \nu_{n-1},1}(u)
    \prod_{1\leq i<j\leq n}S(u_i,u_j)S(u_i,u_j^{-1})\prod_{i=1}^{n-1} K(u_i) \right) = 0.
\end{multline*}

\subsection{Initial condition}
Let \(I_{\mu,\nu}\) denote the right hand side of \eqref{eq:Ptnu} at \(t=0\). We rewrite \(I_{\mu,\nu}\) in terms of \(F_\nu(u_1\ldots u_n)\) and use the properties established in Section \ref{section: F} to prove
\[
I_{\mu,\nu}=\delta_{\mu,\nu}
\]

\subsubsection{Proof of initial condition}
The expression $I_{\mu,\nu}$ can be written using $F_{\nu}$ as follows
   \begin{multline}
    \label{eq:I F}
		I_{\mu,\nu} = (1-q)^n q^{1/2\sum_{i=1}^n (\mu_i-\nu_i)} \oint \frac{\dd u_1}{2\pi\ii u_1} \cdots \oint \frac{\dd u_n}{2\pi\ii u_n}  \\  \times\sum_{\sigma\in \{1,-1\}^n } \sigma\cdot \left(F_\nu(u) \prod_{1\leq i<j\leq n}S(u_i,u_j^{-1}) \prod_{i=1}^n K(u_i) 
		\right) \\ \times \prod_{1\leq i<j\leq n}
		\frac{1}{S(u_i,u_j)S(u_i,u_j^{-1})}
        \prod_{i=1}^n \frac{-1}{K(u_i)} \phi(\mu_i,u_i^{-1}),
	\end{multline}
where the integrals of \eqref{eq:I F} may be evaluated by successively calculating the residues at the points $u_i=q^{-1/2}$. These residues may be evaluated in any order. The poles at the points $u_i=q^{-1/2}$ are of finite order for $\nu\in\mathbb{W}$ and are contained in the function $F_\nu(u)$.
\begin{lemma}
    \label{lm:I F reduced}
    The integral expression \eqref{eq:I F} can be reduced to
\begin{align}
    \label{eq:I F reduced}
    I_{\mu,\nu} = (-1)^n(1-q)^n q^{1/2\sum_{i=1}^n (\mu_i-\nu_i)} \times \oint \frac{\dd u_1}{2\pi\ii u_1} \cdots \oint \frac{\dd u_n}{2\pi\ii u_n}
    \prod_{1\leq i<j\leq n}\frac{1}{S(u_i,u_j)} \prod_{i=1}^n \phi(\mu_i,u_i^{-1})\, F_\nu(u).
\end{align}

\end{lemma}
\begin{proof}
Let us consider the integration on the right-hand side of \eqref{eq:I F} as an explicit nested $n$-fold contour integral chosen with the integration order $u_n\ldots u_1$. This means that the corresponding contours $C_n \ldots C_1$, are aligned such that they all encircle the point $q^{-1/2}$ and that $C_{i+1}$ is contained inside the interior of $C_{i}$. Note that with this alignment, the points of the contour $q^{-1}C_{i+1}^{-1}$, defined as the positively oriented contour with points $q^{-1}u_{i+1}^{-1}$ for $u_{i+1}$ on $C_{i+1}$, may be close to the points of $C_{i+1}$. So, we also choose that the contour $q^{-1}C_{i+1}^{-1}$ is contained inside $C_i$.

We may integrate each summand in \eqref{eq:I F} independently. Consider a summand corresponding to $\sigma$ which has at least one entry equal to $-1$. Such $\sigma$ is of the form $\sigma=(\ldots -1,1\ldots 1)$, i.e. for some $k\in [1\ldots n]$ the values at positions $k\ldots n$ are fixed such that we have $-1$ at position $k$ and $1$ at all positions $k+1\ldots n$. From \eqref{eq:F factors} and \eqref{eq: S K} it can be seen that
$$
\sigma \cdot F_\nu(\ldots u_k\ldots )K(u_k)=F_\nu(\ldots u_k^{-1}\ldots )K(u_k^{-1})
$$
is regular at $u_k=q^{-1/2}$. However in the given integration order we must take into account the factors 
\begin{align}
    S(u_k^{-1},u_l) = \frac{1-u_k u_l}{1-q u_k u_l} 
    \qquad \text{for }l=k+1\ldots n
\end{align}
which occur in $F_\nu(\ldots u_k^{-1}\ldots )$. After we integrate over $u_l$ this factor produces a singularity at $u_k=q^{-1/2}$. On the other hand we may integrate over $u_k$ first by deforming the contour $C_k$ such that it is contained inside the contour $C_l$. In this case we also need to compute the residues at the simple poles $u_k=q^{-1}u_l^{-1}$. This is the same as computing the residues \eqref{eq:F res} and it gives a function which is regular at $u_l=q^{-1/2}$. This also means that $C_k$ may be deformed to lie inside $q^{-1}C_l^{-1}$ (whose orientation is the same as $C_l$). Thus the integral w.r.t. $u_l$ vanishes. This implies that all terms in \eqref{eq:I F} for which $\sigma$ contains a minus sign can be removed and we obtain 
\eqref{eq:I F reduced}. 
\end{proof}

\begin{lemma}
    We have $I_{\mu,\nu}=\delta_{\mu,\nu}$.
\end{lemma}
\begin{proof}
We prove the statement by showing that 
\begin{align}
    \label{eq:I rec}
I_{\mu,\nu} = \delta_{\mu_n,\nu_n}I_{(\mu_1\ldots \mu_{n-1}),(\nu_1\ldots \nu_{n-1})}
\end{align}
To do this we will focus on the integration w.r.t. the variable $u_n$ in \eqref{eq:I F reduced}. Fix the integration domain in \eqref{eq:I F reduced} as in the proof of Lemma~\ref{lm:I F reduced}. Observe that the integrand now has no poles at the points $u_i=q^{-1}u_j^{-1}$ for $i<j$, since these were due to the factors $S(u_{k}^{-1},u_l)$ for $k\neq l$ which may have only occurred for terms  $\epsilon\neq(+1,\dots, +1)$ in \eqref{eq:I F}. And so, we relax the condition that $q^{-1}C_{i+1}^{-1}$ is contained within $C_i$ since they are no longer associated with singularities. Now, insert the recursive formula for $F_\nu(u)$ \eqref{eq:F rec} into $I_{\mu,\nu}$ \eqref{eq:I F reduced}
\begin{align}\label{eq:I I k}
I_{\mu,\nu} = \sum_{k=1}^n I_{\mu,\nu}^{(k)}    
\end{align}
where 
\begin{multline}
    \label{eq:I k}
I_{\mu,\nu}^{(k)} := (-1)^n(1-q)^n q^{1/2\sum_{i=1}^n (\mu_i-\nu_i)} \oint \frac{\dd u_1}{2\pi\ii u_1} \cdots \oint \frac{\dd u_{n-1}}{2\pi\ii u_{n-1}} \\ \prod_{1\leq i<j\leq n-1}\frac{1}{S(u_i,u_j)} \prod_{i=1}^{n-1} \phi(\mu_i,u_i^{-1}) \sum_{\substack{A\subseteq [1\ldots  n-1]\\|A|=k-1}} \prod_{\substack{i\in A\\j\in A^c}}S(u_i,u_j)\,
     F_{\nu_1\ldots \nu_{k-1}}(u_A)F_{\nu_{k+1}\ldots \nu_{n}}(u_{A^c})
     \\ \times\oint \frac{\dd u_n}{2\pi\ii u_n} \phi(\nu_k,u_{n})\phi(\mu_n,u_{n}^{-1})
         \prod_{i\in A^c}\frac{S(u_{n},u_{i})}{S(u_{i},u_{n})}.
\end{multline}
The term with $k=n$ can be calculated and gives
\begin{align}
    \label{eq:I F rec 2}
I_{\mu,\nu}^{(n)} =-(1-q)I_{(\mu_1\ldots \mu_{n-1}),(\nu_1\ldots \nu_{n-1})} \oint \frac{\dd u_n}{2\pi\ii u_n} \phi(\nu_n,u_{n})\phi(\mu_n,u^{-1}_{n}) =\delta_{\mu_n,\nu_n}I_{(\mu_1\ldots \mu_{n-1}),(\nu_1\ldots \nu_{n-1})}.
\end{align}
In order to complete the proof it remains to show that $I_{\mu,\nu}^{(k)}=0$ for $k<n$.  In the second line of \eqref{eq:I k} we have an integral w.r.t. $u_n$ which reads
\begin{align}
    \label{eq:integral u_n}
\oint \frac{\dd u_n}{2\pi\ii u_n} \phi(\nu_k,u_{n})\phi(\mu_n,u_{n}^{-1})
         \prod_{i\in A^c}\frac{S(u_{n},u_{i})}{S(u_{i},u_{n})}
         =
\oint \frac{\dd u_n}{2\pi\ii} 
\frac{-(q^{1/2}-u_n)^{\nu_k-\mu_n-1}}{(1-q^{1/2}u_n)^{\nu_k-\mu_n+1}}
         \prod_{i\in A^c}\frac{u_i-q u_n}{q u_i-u_n}
\end{align}
This integral may only be non-vanishing if $\nu_k\geq \mu_n$. We consider two cases $\nu_k= \mu_n$ and $\nu_k> \mu_n$ separately. 

Fix $k<n$ and consider $I_{\mu,\nu}^{(k)}$ where $\nu_k=\mu_n$. We  integrate each summand in \eqref{eq:I k} with a fixed $A$ independently by first computing the residue of any $u_i$ at $q^{-1/2}$ with $i\in A^c$. Such a change of the integration order will not result in any additional poles for $u_i$. Since there is no singularity at $u_i=q^{-1}u_n^{-1}$, these additional poles may arise due to the factors of the form  $S(u_i,u_j)^{-1}\propto (u_i-u_j)^{-1}$ (with $1\leq i<j\leq n-1$) but these factors are canceled by the terms $S(u_i,u_j)$ (with $i\in A,j\in A^c$) in the integrand \eqref{eq:I k} and the zeroes of $F_{\nu_{k+1}\ldots \nu_n}(u_{A^c})$ (see \eqref{eq:F factors}).  The behavior of the integrand around $u_i=q^{-1/2}$ for $i\in A^c$ is governed by the factors 
\begin{align}\label{eq:F - phi regular}
    \prod_{i\in A^c}\phi(\mu_i,u_i^{-1}) 
     F_{\nu_{k+1}\ldots \nu_{n}}(u_{A^c})
\end{align}
After inserting the explicit form of $\phi$, given in \eqref{eq:phi lambda}, and $F_{\nu_{k+1}\ldots \nu_{n}}(u_{A^c})$, in the form \eqref{eq:F factors}, into \eqref{eq:F - phi regular} and taking into account the particle ordering
$$
\mu_1>\cdots >\mu_n=\nu_{k}>\nu_{k+1}>\cdots >\nu_n
$$
we conclude that \eqref{eq:F - phi regular} is regular at $u_i=q^{-1/2}$ for $i\in A^c$. Hence $I_{\mu,\nu}^{(k)}$ vanishes. 

Next we consider the case $\nu_k> \mu_n$. The integral \eqref{eq:integral u_n} has a pole at $u_n=q^{-1/2}$ of degree $\nu_k-\mu_n+1$. We evaluate this integral by computing the residues which are outside of the integration contour. There is no residue at infinity since the integrand behaves like $u_n^{-2}$ for large $u_n$, so these residues are due to the simple poles at $u_n=q  u_i$ and the evaluation of the integral \eqref{eq:integral u_n} reads
\begin{multline}
    \label{eq:I u_n computed}
\oint \frac{\dd u_n}{2\pi\ii u_n} \phi(\nu_k,u_{n})\phi(\mu_n,u_{n}^{-1})
         \prod_{i\in A^c}\frac{S(u_{n},u_{i})}{S(u_{i},u_{n})}
         \\ =-(1-q^2)q^{-n+k+1+1/2(\nu_k-\mu_n-1)}
         \sum_{l\in A^c}
\frac{u_l (1-q^{1/2} u_l)^{\nu_k-\mu_n-1}}{(1-q^{3/2}u_l)^{\nu_k-\mu_n+1}}
         \prod_{\substack{i\in A^c\\i\neq l}}\frac{u_i-q^2 u_l}{u_i-u_l}.
\end{multline}
We insert this result into $I^{(k)}_{\mu,\nu}$ \eqref{eq:I k} and integrate each term $l\in A^c$ w.r.t. the variable $u_l$. In order to compute the integral over $u_l$ we need to consider the behavior of the integrand around the point $u_l=q^{-1/2}$ as well as $u_l=u_i$ for $i\in A^c$ (due to the change of the order of integration). The latter points are due to the last factor in \eqref{eq:I u_n computed} and $S(u_i,u_j)^{-1}$ appearing in the first line of \eqref{eq:I k}. However, since $F_{\nu_{k+1}\ldots \nu_n}(u_{A^c})$ has a second order zero at $u_l=u_i$ it follows that the integrand is regular at $u_l=u_i$ for $i\in A^c$. The behavior of the integrand around the point $u_l=q^{-1/2}$ is governed by the factors
\[
\phi(\mu_l,u_l^{-1})F_{\nu_{k+1}\ldots \nu_n}(u_{A^c})(1-q^{1/2}u_l)^{\nu_k-\mu_n-1} \propto \frac{1}{(1-q^{1/2}u_l)^{(\nu_{k+1}-\nu_k)+(\mu_n-\mu_l)+2}}.
\]
This can be checked by using the explicit form of $\phi$ and $F$ in the form \eqref{eq:F factors}. Because of the particle ordering we have $\mu_n<\mu_l$ and $\nu_{k+1}<\nu_k$. This means that the integrand is regular at $u_l=q^{-1/2}$ and therefore $I_{\mu,\nu}^{(k)}$ vanishes. 
\end{proof}

\section{Proof of Proposition \ref{prop:Pt_nparticles}: Particle number conservation probability}\label{proof of n-particle probability}

In order to show that $C_t(\mu)$ is given by the integral formula \eqref{integral formula for C} we start with the equation
\begin{align}\label{eq:C}
C_t(\mu)=  \sum_{\nu\in \mathbb W_n}
\mathbb{P}_t(\mu\to \nu),
\end{align}
where we use the integral formula for $\mathbb{P}_t(\mu\to \nu)$ with the integrand expressed in terms of the function $F_\nu(u)$. For $\epsilon\in\{-1,1\}^n$ we define the following quantity
\begin{multline}
    \label{eq:C eps}
    C_t(\mu;\epsilon):=(-1)^n(1-q)^n\\ \times
    \sum_{\nu\in \mathbb W_n}\oint \frac{\dd u_1}{2\pi\ii u_1} \cdots \oint \frac{\dd u_n}{2\pi\ii u_n}
    q^{1/2\sum_{i=1}^n (\mu_i-\nu_i)} \e^{t\left(\sum_{i=1}^n\lambda(u_i)-\alpha\right)} \prod_{i=1}^n \frac{K(u_i^{\epsilon_i})}{K(u_i)} \phi(\mu_i,u_i^{-1})\\ \times
    \prod_{1\leq i<j\leq n}
		\frac{S(u_i^{\epsilon_i},u_j^{-\epsilon_j})}{S(u_i,u_j)S(u_i,u_j^{-1})}
        \,F_\nu(u_1^{\epsilon_1}\ldots u_n^{\epsilon_n}),
\end{multline}
where the integration contours encircle $u_i=q^{-1/2}$ and we choose to integrate in the order of increasing $i$. We denote these contours by $C_i^-$.
\begin{figure}[ht]

\centering

\begin{tikzpicture}[scale=1.1, >=stealth]

\draw[->] (-0.35,0) -- (7.2,0) node[right] {$\Re u$};
\draw[->] (0,-2.3) -- (0,2.3) node[above] {$\Im u$};

\fill (1.2,0) circle (1.5pt)
    node[below=5pt] {$q^{1/2}$};

\fill (3.0,0) circle (1.5pt)
    node[below=5pt] {$1$};

\fill (5.3,0) circle (1.5pt)
    node[below=5pt] {$q^{-1/2}$};

\draw[thick,postaction={decorate},
decoration={markings,mark=at position 0.2 with {\arrow{>}}}]
(5.3,0) circle (0.75);
\draw[thick,postaction={decorate},
decoration={markings,mark=at position 0.2 with {\arrow{>}}}]
(5.3,0) circle (1.45);

\node at (5.3,0.98) {$C_1^-$};
\node at (5.3,1.68) {$C_n^-$};

\fill (6.21,0.29) circle (0.6pt);
\fill (6.35,0.38)  circle (0.6pt);
\fill (6.49,0.46)     circle (0.6pt);
\end{tikzpicture}

\caption{Nested contours around $q^{-1/2}$.}

\label{fig:nested-contours}

\end{figure}
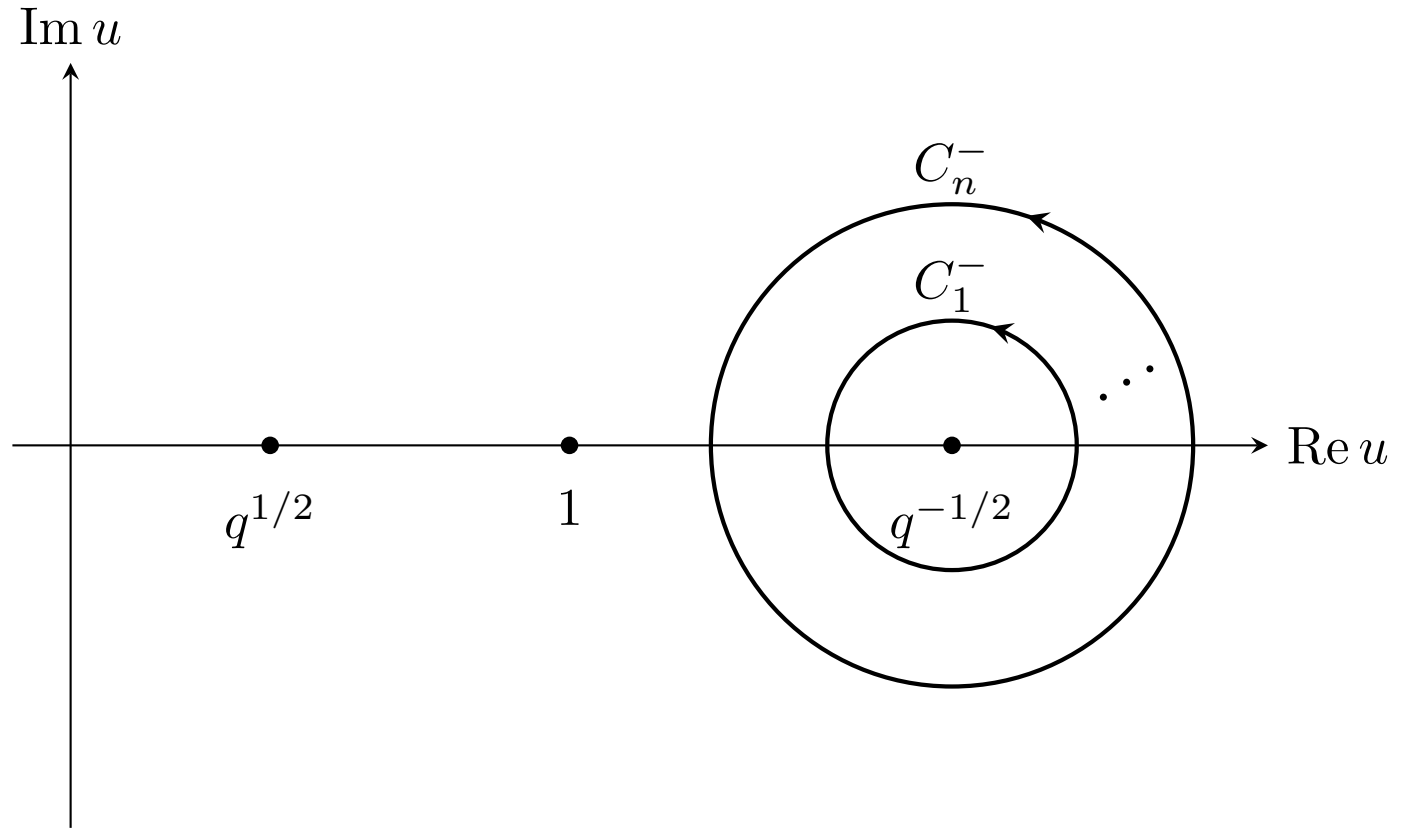
Then $C_t(\mu)$ is given by 
\begin{align}\label{eq:C C eps}
C_t(\mu)=\sum_{\epsilon\in\{-1,1\}^n} C_t(\mu;\epsilon)
\end{align}
We will focus on the computation of the individual terms $C_t(\mu;\epsilon)$ and after that we will insert the result in \eqref{eq:C C eps}. In the calculation below the contours of the initial multiple integral encircle $q^{-1/2}$; to perform the sum over final configurations they are replaced by sign-dependent contours and after this sum has been evaluated these contours are deformed to common contours encircling both $q^{-1/2}$ and $0$ before the sum over signs is performed. This computation is similar in spirit to a computation of ASEP particle position formulas from \cite{tracy_intformula_2008}.

 \subsection{Calculation of $C_t(\mu;\epsilon)$}
To calculate $C_t(\mu;\epsilon)$ we must interchange the sum over $\nu\in\mathbb W_n$ with the contour integrals. Proposition~\ref{prop: sum nu} permits this under the condition
\begin{align}\label{eq:convergence condition x}
        \left|\frac{1 -q^{-1/2} u_k^{\epsilon_k}}{1-q^{1/2} u_k^{\epsilon_k}} \right|<1\qquad \text{for } k=1\ldots n
\end{align}
The contours in the expression of $C_t(\mu;\epsilon)$ go around the point $q^{-1/2}$ which means that for $k$ such that $\epsilon_k=-1$ this condition is satisfied in the integration contour. However, for $k$ with $\epsilon_k=1$ we have 
\begin{align}\label{eq:convergence condition +1}
        \left|\frac{1 -q^{-1/2} u_k}{1-q^{1/2} u_k} \right|<1
\end{align}
which cannot be satisfied in the integration contour. Therefore we must modify the integration contours depending on $\epsilon_k$. 

We encode $\epsilon\in\{-1,1\}^n$ by the subset $A\subseteq[1\ldots n]$ of positions where $\epsilon_i=-1$. Thus $A^c$ records the positions where $\epsilon_i=1$. For fixed $A$ choose the contours as follows:
\begin{align}
\label{eq:choice C^-}
&u_i \quad \text{is integrated over contour $C_i^-$ that encircles } q^{-1/2} \quad \text{if }i\in A
\\
\label{eq:choice C^+}
&u_i \quad \text{is integrated over contour $C_i^+$ that {\bf{does not}} encircle } q^{-1/2} \quad \text{if }i\in A^c
\end{align}
By a direct inspection of the integrand of \eqref{eq:C eps} one can verify using \eqref{eq:F} that there is no residue at infinity for the integration variables $u_i$ with $i\in A^c$. Each $C_i^+$ thus is a collection of small clockwise oriented contours around singularities of the integrand which we need to examine. For $i\in A^c$ we write $u_i\sim x$ when $C_i^+$ contains a small contour around $x\in\mathbb C$. To list these singularities we rewrite the second line of \eqref{eq:C eps} using $A$ and $A^c$:
\begin{multline*}
     \prod_{1\leq i<j\leq n}
		\frac{S(u_i^{\epsilon_i},u_j^{-\epsilon_j})}{S(u_i,u_j)S(u_i,u_j^{-1})}
        F_\nu(u^\epsilon) = \\
    \prod_{i\in A}\frac{1}{(q^{1/2}-u_i^{-1})(1-q^{1/2} u_i^{-1})^{\nu_1-1}} 
    \prod_{i\in A^c}\frac{1}{(q^{1/2}-u_i)(1-q^{1/2} u_i)^{\nu_1-1}} \nonumber
    \\
    \times
    \prod_{\substack{i,j\in A \\ j< i}} \frac{(u_i-u_j)(q -u_i u_j)}{(u_i-q u_j)(1-q u_i u_j)}
    \prod_{\substack{i,j\in A^c \\ j< i}} \frac{u_i-u_j}{u_i-q u_j} 
    \prod_{\substack{i\in A \\ j\in A^c}} \frac{1-u_i u_j}{1-q u_i u_j} \prod_{\substack{i\in A \\ j\in A^c\\ i<j}} \frac{u_i-q u_j}{u_j-q u_i}\times \widetilde F_\nu(u_A,u_{A^c}),
\end{multline*}
where $\widetilde F_\nu$ is defined by \eqref{eq:F factors}. 
As a result the contours of $C_i^+$ are described by a collection of small contours given by
\begin{align}
    \label{eq:poles 0}
u_i\sim q^{1/2}
\qquad 
&\text{for } i\in A^c 
\end{align}
and potentially
\begin{align}
    \label{eq:poles 1}
    &u_i\sim  q u_j 
    &\quad 
    &\text{ for } j<i \text{ and } i,j\in A^c 
    \\
    \label{eq:poles 2}
    &u_i \sim q^{-1} u_j 
    &\quad 
    &\text{ for } j>i \text{ and } i,j\in A^c 
    \\
    \label{eq:poles 3}
    &u_i\sim  q^{-1} u_j^{-1} 
    &\quad 
    &\text{ for } i\in A^c, \, j\in A 
    \\
    \label{eq:poles 4}
    &u_i\sim q u_j 
    &\quad 
    &\text{ for } j>i \text{ and } i\in A^c, \, j\in A
\end{align}
if these regions enclose any poles. 
\begin{prop}
    Fix $\epsilon=(\epsilon_1\ldots \epsilon_n )\in\{-1,1\}^n$ and let $D_\epsilon$ denote the region enclosed by  $C_1^{\epsilon_1}, \ldots, C_n^{\epsilon_n}$. The choice of the contours \eqref{eq:choice C^-} and \eqref{eq:choice C^+} in the integral \eqref{eq:C eps} is such that the conditions \eqref{eq:convergence condition x} hold in the region $D_{\epsilon}$.
\end{prop}
\begin{proof}
It is clear that for $n=1$ and $0<q<1$ the two conditions 
\begin{align}
\label{eq:convergence condition pm}
        \left|\frac{1 -q^{-1/2} u_k^{\pm 1}}{1-q^{1/2} u_k^{\pm 1}} \right|<1
\end{align}
with $k=1$, hold in the respective integration domains given by \eqref{eq:choice C^+} and \eqref{eq:choice C^-}. Hence the statement is true for $D_{(\epsilon_1)}$. Suppose the Proposition holds in the case of $n-1$-fold integral. Then \eqref{eq:convergence condition pm}  holds for $k=2\ldots n$ in the domain $D_{(\epsilon_2\ldots \epsilon_n)}$ given by $n-1$ contours $C_2^{\epsilon_2}, \ldots, C_n^{\epsilon_n}$.
We will complete the proof by checking that \eqref{eq:convergence condition pm} also holds in the two cases
\begin{align}
    \label{eq:integration domain 1-}
&D_{(-1,\epsilon_2,\ldots \epsilon_n)},\qquad \epsilon_1=-1\\
\label{eq:integration domain 1+}
&D_{(1,\epsilon_2,\ldots \epsilon_n)},\qquad \epsilon_1=1
\end{align}
We need to check that after combining $C_1^\pm$ with the existing contours $C_2^{\epsilon_2}, \ldots, C_n^{\epsilon_n}$ the convergence condition for the variable $u_1$ holds.

In the following we will use the labeling by subsets $A$ and $A^c$ that corresponds to the fixed $\epsilon$. 
First we consider the case $\epsilon_1=-1$. The contour \(C_1^-\) can be chosen sufficiently close to \(q^{-1/2}\) and the convergence condition holds. Next we consider the case \eqref{eq:integration domain 1+}. First we check the domain enclosed by
\begin{align}
    \label{eq:proof convergence C+C+}
C_1^+ \times C_m^+
\end{align}
Since $1\in A^c$ and $m\in A^c$ we need to consider a possible pole in the domain \eqref{eq:poles 2} where $u_1\sim  q^{-1} u_m$, for $m>1$. Since the contour of $u_m$ encircles $q^{1/2}$ we get 
\begin{align}\label{eq:proof convergence C-C+ Y}
(u_1\sim q^{-1/2}) \times (u_m\sim q^{1/2})
\end{align}
The inclusion of this domain in the integration would violate the condition \eqref{eq:convergence condition pm} for $k=1$. However, the function $\widetilde F_\nu(u_1\ldots u_m=q^{1/2}\ldots)$ has a divisor $(1-q^{1/2}u_1)$ due to Proposition \ref{prop: wheel 2}. Hence \eqref{eq:proof convergence C-C+ Y} can be excluded from the domain \eqref{eq:proof convergence C+C+}. 

Finally we check the domain \begin{align}
    \label{eq:proof convergence C+C-}
C_1^+ \times C_m^-
\end{align}
In this case $1\in A^c$ and $m\in A $ and we need to consider possible poles in the domains \eqref{eq:poles 3} and \eqref{eq:poles 4}.  The domain
\eqref{eq:poles 4} corresponds to $u_1\sim  q u_m$ for $m>1$ and since $u_m$ encircles $q^{-1/2}$ the condition \eqref{eq:convergence condition x} holds for $u_1$. The domain of \eqref{eq:poles 3} corresponds to $u_1\sim q^{-1}u_m^{-1}$ which leads to
\begin{align}\label{eq:proof convergence C-C+ Z}
(u_1\sim q^{-1/2}) \times (u_m\sim q^{-1/2})
\end{align}
In this domain the function $\widetilde F_\nu(u_1\ldots u_m^{-1}=q^{1/2}\ldots)$ has a divisor $(1-q^{1/2}u_1)$ due to Proposition \ref{prop: wheel 2}. Hence there is no pole inside \eqref{eq:proof convergence C-C+ Z} and thus we excluded it from the integration domain \eqref{eq:proof convergence C+C-}. This analysis completes the proof.
\end{proof}

The proposition implies that the convergence conditions \eqref{eq:convergence condition x} hold uniformly on the integration contours and justifies interchanging the sum and integrals. We now use \eqref{eq:F summed} to perform the sum over $\nu\in\mathbb W_n$:
\begin{multline}
    \label{eq:C eps summed}
    C_t(\mu;\epsilon)=(-1)^{|A^c|}
    \oint_{C_1^{\epsilon_1}} \frac{\dd u_1}{2\pi\ii u_1} \cdots \oint_{C_n^{\epsilon_n}} \frac{\dd u_n}{2\pi\ii u_n}
    q^{1/2\sum_{i=1}^n \mu_i} \e^{t\left(\sum_{i=1}^n\lambda(u_i)-\alpha\right)} \prod_{i=1}^n \frac{K(u_i^{\epsilon_i})}{K(u_i)} \phi(\mu_i,u_i^{-1})\\
    \prod_{i=1}^{n}\frac{(1-q^{1/2} u_i^{\epsilon_i})}{(q^{1/2}- u_i^{\epsilon_i})u_i^{\epsilon_i}}
    \prod_{1\leq i<j\leq n}
		\frac{S(u_i^{\epsilon_i},u_j^{-\epsilon_j})S(u_i^{\epsilon_i},u_j^{\epsilon_j})S(u_j^{\epsilon_j},u_i^{\epsilon_i})}{S(u_i,u_j)S(u_i,u_j^{-1})}.
\end{multline}

\subsection{Calculation of $C_t(\mu)$}
In this section we compute the sum of $C_t(\mu;\epsilon)$ over $\epsilon\in \{1,-1\}^n$ and finalize the proof of Proposition \ref{prop:Pt_nparticles}. We first match the integration contours of $C_t(\mu;\epsilon)$ and then compute the sum over $\epsilon\in \{1,-1\}^n$ using a summation identity. 

We modify the contours of $C_t(\mu;\epsilon)$ given by \eqref{eq:C eps summed} such that they do not depend on $\epsilon$. The integrand in \eqref{eq:C eps summed} has an additional pole at $u_i=0$ for $i\in A^c$ (i.e. $\epsilon_i=1$) compared to the integrand in \eqref{eq:C eps}. We can deform all contours $C_i^+$ and integrate over the singularities outside of $C_i^+$. This results in an integral over a contour which surrounds $q^{-1/2}$ and $0$, hence
\begin{multline}
    \label{eq:Ceps}
    C_t(\mu;\epsilon)=
    \oint \frac{\dd u_1}{2\pi\ii u_1} \cdots \oint \frac{\dd u_n}{2\pi\ii u_n}
    q^{1/2\sum_{i=1}^n \mu_i} \e^{t\left(\sum_{i=1}^n\lambda(u_i)-\alpha\right)} \prod_{i=1}^n  \phi(\mu_i,u_i^{-1})\\
    \prod_{i=1}^{n}\frac{K(u_i^{\epsilon_i})}{K(u_i)}\frac{(1-q^{1/2} u_i^{\epsilon_i})}{(q^{1/2}- u_i^{\epsilon_i})u_i^{\epsilon_i}}
    \prod_{1\leq i<j\leq n}
		\frac{S(u_i^{\epsilon_i},u_j^{-\epsilon_j})S(u_i^{\epsilon_i},u_j^{\epsilon_j})S(u_j^{\epsilon_j},u_i^{\epsilon_i})}{S(u_i,u_j)S(u_i,u_j^{-1})},
\end{multline}
where all contours encircle $q^{-1/2}$ and $0$ and no other singularities and are oriented counterclockwise.

In the next step we compute
\begin{align}\label{eq:C C eps X}
C_t(\mu)=\sum_{\epsilon\in\{-1,1\}^n} C_t(\mu;\epsilon)
\end{align}
It is convenient to use the labelling by subsets $A$ instead of $\epsilon$. We can write the product over $i$ and the double product over $i,j$ appearing in the second line of \eqref{eq:Ceps} using the subsets $A,A^c\subseteq [1\ldots n]$ as follows
\begin{align}
    \label{eq:a subsets product i}
\prod_{i=1}^{n}\frac{K(u_i^{\epsilon_i})}{K(u_i)}\frac{(1-q^{1/2} u_i^{\epsilon_i})}{(q^{1/2}- u_i^{\epsilon_i})u_i^{\epsilon_i}}&=
\frac{(-1)^n(1-q)^n}{(1-a q^{1/2})^n}
\prod_{i=1}^n \frac{K(u_i^{-1})}{(1-u_i^2)}
\prod_{i\in A} (1-a u_i)u_i
\prod_{i\in A^c} (a-u_i),
\end{align}
\begin{multline}
    \label{eq:a subsets product ij}
    \prod_{1\leq i<j\leq n}
		\frac{S(u_i^{\epsilon_i},u_j^{-\epsilon_j})S(u_i^{\epsilon_i},u_j^{\epsilon_j})S(u_j^{\epsilon_j},u_i^{\epsilon_i})}{S(u_i,u_j)S(u_i,u_j^{-1})} \\= 
        \prod_{1\leq i<j\leq n}
		S(u_j,u_i)S(u_j^{-1},u_i)
        \times\prod_{\substack{i,j\in A\\ i<j}} \frac{1}{S(u_i,u_j^{-1})}
        \prod_{\substack{i,j\in A^c\\ i<j}} \frac{1}{S(u_i^{-1},u_j)}
        \prod_{\substack{i\in A\\ j\in A^c}} \frac{1}{S(u_i,u_j)}.
\end{multline}
Using these two identities in \eqref{eq:Ceps} and substituting the result into \eqref{eq:C C eps X} gives us
\begin{multline}
    \label{eq:C subsets}
C_t(\mu)=
\frac{(-1)^n(1-q)^n}{(1-a q^{1/2})^n} \\ \times
\oint \frac{\dd u_1}{2\pi\ii u_1} \cdots \oint \frac{\dd u_n}{2\pi\ii u_n}
    q^{1/2\sum_{i=1}^n \mu_i} \e^{t\left(\sum_{i=1}^n\lambda(u_i)-\alpha\right)}  
    \prod_{i=1}^n \phi(\mu_i,u_i^{-1})
    \frac{K(u_i^{-1})}{(1-u_i^2)}
    \prod_{1\leq i<j\leq n}
		S(u_j,u_i)S(u_j^{-1},u_i)\\ \times
        \sum_{A\subseteq [1\ldots n]}
    \left(
    \prod_{i\in A} (1-a u_i)u_i
    \prod_{\substack{i,j\in A\\ i<j}} \frac{1}{S(u_i,u_j^{-1})}\right)
    \left(\prod_{i\in A^c} (a-u_i)
        \prod_{\substack{i,j\in A^c\\ i<j}} \frac{1}{S(u_i^{-1},u_j)}\right)
        \prod_{\substack{i\in A\\ j\in A^c}} \frac{1}{S(u_i,u_j)}.
\end{multline}
We calculate the summation in the third line of \eqref{eq:C subsets} in the following Lemma
\begin{lemma}
    We have the identity
    \begin{multline}
        \label{eq:subsets sum}
        \sum_{A\subseteq [1\ldots n]}
    \left(
    \prod_{i\in A} (1-a u_i)u_i
    \prod_{\substack{i,j\in A\\ i<j}} \frac{1}{S(u_i,u_j^{-1})}\right)\\ \times
    \left(\prod_{i\in A^c} (a-u_i)
        \prod_{\substack{i,j\in A^c\\ i<j}} \frac{1}{S(u_i^{-1},u_j)}\right)
        \prod_{\substack{i\in A\\ j\in A^c}} \frac{1}{S(u_i,u_j)}= a^n \prod_{i=1}^n(1-u_i^2).
    \end{multline}
\end{lemma}
\begin{proof}
    Denote the left-hand side of \eqref{eq:subsets sum} by
    $H_n(a;u_1 \ldots u_n)$ and substitute $S(u,v)$
    \begin{multline}
        \label{eq: H_n}
      H_n(a;u_1 \ldots u_n):=  \sum_{A\subseteq [1\ldots n]}
    \left(
    \prod_{i\in A} (1-a u_i)u_i
    \prod_{\substack{i,j\in A\\ i<j}} \frac{q- u_i u_j}{1-u_i u_j}\right) \\ \times
    \left(\prod_{i\in A^c} (a-u_i)
        \prod_{\substack{i,j\in A^c\\ i<j}} \frac{1- q u_i u_j}{1-u_i u_j}\right)
        \prod_{\substack{i\in A\\ j\in A^c}} \frac{u_i-q u_j}{u_i-u_j}.
    \end{multline}       
    We need to prove that 
        \begin{align}
        \label{eq: H_n summed}
H_n(a;u_1 \ldots u_n)= a^n \prod_{i=1}^n(1-u_i^2).
    \end{align}
    We prove this identity by induction on $n$.
    The identity \eqref{eq: H_n summed} holds for $n=1$
    \[
        H_1(a;u_1)=(1-a u_1)u_1+(a-u_1)
        =a(1-u_1^2)
    \]
We first show that $H_n$ is symmetric in $u_1 \ldots u_n$. Consider swapping $u_k$ with $u_l$ in individual terms in $H_n$. If $k,l\in A$ or $k,l \in A^c$ then the corresponding terms in \eqref{eq: H_n} are manifestly symmetric. The two terms where $k\in A,~l\in A^c$ and $k\in A^c,~l\in A$ get exchanged after swapping $u_k$ with $u_l$ hence their sum is invariant. Therefore $H_n$ is symmetric in $u_1 \ldots u_n$. 

For fixed $u_1 \ldots u_n$ the function $H_n$ is a polynomial in $a$ of degree at most $n$. We determine this polynomial at $n+1$ points. By the symmetry it is enough first to specialise $a=u_n$. In this specialisation every summand with $n\in A^c$ vanishes. In the
    surviving terms, we identify $A\setminus\{n\}$ with a subset $A$ of
    $\{1 \ldots n-1\}$. The factors involving $u_n$ then combine with the
    one-variable factors as
    \begin{align*}
        (1-u_nu_i)u_i\frac{q-u_i u_n}{1-u_i u_n}
            &=q\left(1-\frac{u_n}{q}u_i\right)u_i
            &\quad &\text{for } i\in A\\
        (u_n-u_i)\frac{u_n-q u_i}{u_n-u_i}
            &=q\left(\frac{u_n}{q}-u_i\right)
            &\quad &\text{for } i\in A^c
    \end{align*}
    Therefore we have
    \begin{equation}
        \label{eq:subset induction at u}
        H_n(u_n;u_1 \ldots u_n)
        =u_n(1-u_n^2)q^{n-1}
          H_{n-1}\left(\frac{u_n}{q};u_1 \ldots u_{n-1}\right)
    \end{equation}
The induction hypothesis allows us to use \eqref{eq: H_n summed} on the right hand side which gives 
    \[
        H_n(u_n;u_1 \ldots u_n)
        =u_n^n\prod_{i=1}^n(1-u_i^2)
    \]
    By symmetry the same evaluation holds at $a=u_k$ for every
    $k=1\ldots n$.

We use $a=u_n^{-1}$ as the remaining interpolation point. Now every summand with $n\in A$ vanishes. In the surviving terms, we identify $A$ with a subset of $\{1 \ldots n-1\}$. The factors involving $u_n$ then combine with the one-variable factors as
    \begin{align*}
        \left(1-\frac{u_i}{u_n}\right)u_i
        \frac{u_i-q u_n}{u_i-u_n}
            &=q\left(1-\frac{u_i}{q u_n}\right)u_i
            &\quad&\text{for } i\in A\\
        \left(\frac{1}{u_n}-u_i\right)
        \frac{1-q u_i u_n}{1-u_i u_n}
            &=q\left(\frac{1}{q u_n}-u_i\right)
            &\quad& \text{for } i\in A^c
    \end{align*}
Hence
    \begin{equation}
        \label{eq:subset induction at inverse u}
        H_n(u_n^{-1};u_1 \ldots u_n)
        =\frac{1-u_n^2}{u_n}q^{n-1}
          H_{n-1}\left(\frac{1}{q u_n};u_1 \ldots u_{n-1}\right)
    \end{equation}
    and the induction hypothesis gives us
    \[
        H_n(u_n^{-1};u_1 \ldots u_n)
        =u_n^{-n}\prod_{i=1}^n(1-u_i^2)
    \]
    Thus the two polynomials in $a$ appearing in \eqref{eq: H_n summed}, which are
    both of degree at most $n$, agree at
    $a=u_1 \ldots u_n,u_n^{-1}$. For generic values of the $u_i$ these are
    $n+1$ distinct points so the polynomials are identical. This proves
    \eqref{eq:subsets sum} and hence the lemma.
\end{proof}

Substituting \eqref{eq:subsets sum} into \eqref{eq:C subsets} cancels the factors $(1-u_i^2)^{-1}$ in the first line of \eqref{eq:C subsets}. Using the definition of $a$ in the resulting formula gives us
\begin{multline}
    \label{eq:C computed}
C_t(\mu)=\alpha^n q^{1/2(\sum_{i=1}^n \mu_i-n)} 
\oint \frac{\dd u_1}{2\pi\ii u_1} \cdots \oint \frac{\dd u_n}{2\pi\ii u_n}
\e^{t\left(\sum_{i=1}^n\lambda(u_i)-\alpha\right)}   \\ \times
    \prod_{1\leq i<j\leq n}
		S(u_j,u_i)S(u_j^{-1},u_i)    \prod_{i=1}^n K(u_i^{-1}) \phi(\mu_i,u_i^{-1}).
\end{multline}
This completes the proof of Proposition \ref{prop:Pt_nparticles}.

\bibliographystyle{iopart-num-long}
\bibliography{Library_new}{}

\end{document}